# The Quantum Dynamics of Cost Accounting: Investigating WIP via the Time-Independent Schrödinger Equation

Maksym Lazirko
Department of Accounting & Information Systems , Rutgers University

*For all inquiries , please email [mol12@business.rutgers.edu](mol12@business.rutgers.edu) (Maksym Lazirko)*
*Rutgers Accounting Research Center , 1 Washington Pl, Newark, NJ 07102*

**Abstract:**

The intersection of quantum theory and accounting presents a novel and intriguing frontier in exploring financial valuation and accounting practices. This paper applies quantum theory to cost accounting, specifically Work in Progress (WIP) valuation. WIP is conceptualized as materials in a quantum superposition state whose financial value remains uncertain until observed or measured. This work comprehensively reviews the seminal works that explored the overlap between quantum theory and accounting. The primary contribution of this work is a more nuanced understanding of the uncertainties involved, which emerges by applying quantum phenomena to model the complexities and uncertainties inherent in managerial accounting. In contrast, previous works focus more on financial accounting or general accountancy.

**Introduction:**

The traditional accounting treatment of Work in Progress (WIP) involves estimating the value of partially completed goods, which poses inherent challenges due to the production processes' uncertain and dynamic nature (Budd, 2010). By drawing parallels with quantum superposition, where particles exist in multiple states until measured(Frino, 2015), we can propose a novel perspective on accounting for WIP. This quantum-inspired approach addresses the intrinsic uncertainties in valuing WIP and aligns with the evolving landscape of accounting standards that increasingly recognize the complexity of modern business operations.

To effectively study the application of quantum theory to managerial accounting, we integrate principles from quantum mechanics, such as superposition and entanglement, with accounting theories related to asset valuation and measurement. This interdisciplinary approach will involve a theoretical exploration supported by conceptual analysis and examples from quantum physics and accounting practices. By examining the parallels between quantum phenomena and the valuation challenges of WIP, we aim to develop a robust framework that can offer new insights into accounting methodologies.

This endeavor aims to furnish a thorough examination of conventional accounting methodologies concerning WIP and elucidate the intricacies inherent in the valuation processes associated with such practices. The literature review will expound upon the traditional accounting paradigms employed in WIP valuation, expatiating on the challenges that pervade these established frameworks. Subsequently, the paper will pivot to an unconventional perspective by introducing key tenets from quantum theory, specifically delving into the quantum phenomena of superposition and entanglement. This section aims to elucidate the nuanced connections between quantum concepts and their potential applicability to accounting. Following this quantum theory exposition, the research will articulate a robust conceptual framework synthesizing quantum principles with accounting practices. It will focus on their integration to mitigate valuation uncertainties inherent in WIP accounting. This conceptual development will address the perennial challenges of traditional accounting methodologies. The subsequent segment will scrutinize the implications of adopting this quantum-inspired accounting model for WIP on established accounting standards, practices, and the overarching financial reporting landscape. This exploration will navigate the potential transformative effects on accounting, considering the theoretical underpinnings and practical consequences of embracing quantum principles in financial reporting. Ultimately, the paper aspires to contribute to the ongoing discourse surrounding accounting innovation by integrating quantum theory into the valuation framework of Work-in-Progress, fostering a deeper understanding of the intricate dynamics at play within contemporary financial reporting practices.

Integrating quantum theory into accounting, particularly for WIP, is not merely an academic exercise but a necessary evolution in response to the complexities of modern business environments. Quantum computing maturing as a technology means accountants should keep up with the times (Lazirko, 2023). As production processes become more intricate and interconnected (Kovács, 2016), traditional accounting methods struggle to accurately capture the dynamic and uncertain nature of WIP valuation (Myrelid & Olhager, 2015). By adopting a quantum-inspired approach, accountants can leverage a more flexible and nuanced framework that reflects the realities of contemporary business operations (J. S. Demski et al., 2006). This paper aims to demonstrate that quantum theory offers valuable insights and tools to enrich accounting practices, making a compelling case for accountants to explore and embrace this innovative perspective.

**Literature Review**

The following literature review summarizes key findings from several seminal works in quantum accounting. It discusses the current gap in the literature regarding the application of quantum theory to WIP accounting. In accordance with the Preferred Reporting Items for Systematic Reviews and Meta-Analyses (PRISMA) guidelines[1], the literature review section is systematically structured to include the selection criteria, search strategy, data collection and analysis, and study quality and risk of bias. The following rephrased paragraphs are presented to fit within this structured approach:

---
[1] http://www.prisma-statement.org/

### *Selection Criteria*

The inclusion criteria for this review were studies that examined integrating quantum theory with accounting practices. Specifically, the review focused on seminal works that explored the conceptual applications of quantum information to accounting information systems, the valuation of WIP, and the implications of quantum features such as entanglement and superposition in accounting.

### *Search Strategy*

The search strategy involved identifying key studies that have contributed to the field of quantum accounting. The search included works by (De Oliveira & Lustosa, 2022; J. Demski et al., n.d.; J. S. Demski et al., 2006, 2009; J. Fellingham et al., 2022; J. C. Fellingham et al., 2018; J. Fellingham & Schroeder, 2006). These studies are selected for their pioneering exploration of the intersection between quantum information theory and accounting principles. They represent a comprehensive list focusing on the intersection of quantum and accounting.

*Data Collection & Analysis*

Data were extracted from the selected studies to understand the proposed theoretical frameworks and the potential for quantum theory to inform accounting practices. The analysis focused on the conceptual overlap between quantum theory and accounting, applying quantum probabilities to double-entry information processing, and exploring topology and entropy in accounting information systems.

*Study Quality & Risk of Bias*

The quality of the included studies was assessed based on their conceptual rigor and the novelty of the approaches proposed. The risk of bias is considered low, as the studies were primarily theoretical explorations of quantum theory's application to accounting.

*Synthesis of Results*

- Demski et al. (2006) proposed that quantum theory could provide a novel perspective on the fundamental laws of accounting, potentially enriching the approach to accounting information (J. S. Demski et al., 2006).
- Fellingham and Schroeder (2006) introduced the use of quantum probabilities in a two-agent control setting, suggesting that quantum interference could lead to more efficient double-entry information processing (J. Fellingham & Schroeder, 2006).
- Demski et al. (2009) investigated the conceptual applications of topology within quantum information and accounting, highlighting the abstract mathematical concepts underpinning accounting information systems (J. S. Demski et al., 2009).

- In a subsequent study, Demski et al. (2009) explored the role of topological quantum computation in addressing issues of decoherence and imprecision in quantum computation, proposing the use of exotic topological states for global information storage and manipulation (J. S. Demski et al., 2009).
- Fellingham, Lin, and Schroeder (2018) examined quantum entropy and its potential applications in accounting, suggesting parallels between the uncertainty in quantum systems and financial information (J. C. Fellingham et al., 2018).
- Fellingham, Lin, and Schroeder (2022) discussed the interrelation between entropy, double-entry accounting, and quantum entanglement, providing insights into the potential integration of these concepts within accounting frameworks (J. Fellingham et al., 2022).
- De Oliveira and Lustosa (2022) applied quantum physics concepts to the economic nature of goodwill, proposing that the entanglement of intangible and physical capital forms a company's economic value (De Oliveira & Lustosa, 2022).
- Lazirko (2023) compared quantum standards between the European Union and the US while exploring their implications on AIS (Lazirko, 2023).

Despite the intriguing insights provided by the aforementioned works, there remains a significant gap in the literature regarding applying quantum theory to accounting. The concept of WIP as materials in a state of quantum superposition, with uncertain financial value until measured, is a novel approach that has yet to be extensively explored. The existing literature has laid the groundwork by establishing the relevance of quantum concepts such as entanglement, superposition, and entropy in the context of accounting information systems and the valuation of intangible assets like goodwill. However, applying these principles to the valuation and accounting of WIP requires further theoretical development and empirical investigation.

The current literature primarily focuses on the conceptual overlap between quantum theory and accounting principles, emphasizing the potential for quantum theory to provide a new lens through which to view accounting challenges. Applying quantum theory to managerial accounting would necessitate a more detailed exploration of how quantum measurement, uncertainty, and the dynamics of superposition could be analogously applied to the valuation of materials and goods that are in the production process but still need to be completed. This represents an opportunity for this research to bridge the gap by developing a quantum accounting information systems framework.

**Theoretical Framework**

*Introduction to Quantum Concepts Relevant to Accounting*

Integrating quantum concepts into accounting practices offers a novel perspective on addressing the complexities and uncertainties inherent in financial information systems (J. Fellingham & Schroeder, 2006). This section introduces several quantum concepts that are particularly relevant to developing a quantum accounting model for WIP.

<u>Entanglement</u>

Quantum entanglement describes a phenomenon where the state of one particle cannot be described independently of the state of another, regardless of the distance separating them. In accounting, this concept can be applied to understand the interconnectedness of various accounting entities and transactions, where the change in one can instantaneously affect the state of another(Orrell, 2019).

<u>Superposition</u>

Superposition refers to quantum systems' ability to exist simultaneously until measured in multiple states (Lee, 2021). Rather than make an assumption about a system and collapse the unobserved[2] to a single point, we can allow it to occupy multiple states instead. Moreover, accounting systems collapse as a wave function when something happens to justify the collapse. In managerial accounting, superposition can be used to model the uncertain value of WIP. It can embody multiple potential values until an observation (e.g., sale or completion) collapses it to a single value.

---
[2] Exogenous variables as a comprehensive measure are inherently ubovservable because we can not capture all of them.

### Measurement

Quantum measurement is the process by which the state of a quantum system becomes known, collapsing superposition into one of the possible states (Peres, 2000). For accounting, the measurement process can analogously represent the valuation and recognition of financial information, transforming uncertain or probabilistic values into definite figures.

### Teleportation

Quantum teleportation involves the transfer of quantum states from one location to another without the physical movement of the particles(Wang et al., 2015). While more abstract in its application to accounting, teleportation could metaphorically represent the transfer of value or information across different parts of an organization instantaneously, without the traditional transfer of documentation or entries.

### Entropy

Entropy measures the uncertainty or disorder within a quantum system in quantum mechanics. Applied to accounting, entropy can represent the complexity and uncertainty in financial information, highlighting the challenges in achieving accurate and precise measurements of financial states (J. C. Fellingham et al., 2018), such as the valuation of WIP.

Quantum Tunneling

Quantum tunneling, a fundamental concept in quantum mechanics, explains that a particle can tunnel through an energy barrier that it would not be able to surmount classically, a phenomenon that is not directly observable but can be inferred from the presence of the particle on the other side of the barrier (Davis & Heller, 1981). Similarly, arbitrage opportunities can be thought of as "energy barriers" that can be "tunneled" through by investors seeking to profit from price differences in different markets. This analogy suggests that quantum tunneling could be used to model arbitrage behavior, providing insights into the dynamics of markets and the risks associated with arbitrage activities. In accounting research, we historically utilize Bayesian theory for inference problems (Swieringa et al., 1976).

*Overview of WIP in Accounting & Its Challenges*

Work in Progress (WIP) refers to the goods in production that are not yet completed. Accounting for WIP involves estimating the value of these goods, which can be challenging due to the dynamic nature of production processes and the uncertainty regarding the outcome and costs. Traditional accounting methods often struggle to capture real-time fluctuations and inherent uncertainties in value, leading to potential inaccuracies in financial reporting (Almeida et al., 2020).

**Methodology**

This section outlines the methodology for integrating quantum theory with WIP accounting, focusing on applying quantum entanglement, superposition, and measurement theory to model the complexities and uncertainties inherent in WIP valuation. Drawing parallels between the abovementioned quantum concepts and the challenges in accounting for WIP, we can propose a quantum accounting model for WIP that leverages the principles of superposition, entanglement, measurement, teleportation, and entropy. This model conceptualizes WIP as existing in a state of superposition, embodying multiple potential values based on different outcomes of the production process. The value of WIP is entangled with other financial entities and transactions, reflecting the interconnected nature of business operations.

The model suggests that the value of WIP can exist in multiple potential states (superposition) until an event (e.g., completion of production, sale) leads to the measurement and recognition of a specific value. This approach acknowledges the uncertainty in WIP valuation. It provides a framework for dynamically updating the estimated value as new information becomes available that aligns with current practices such as continuous monitoring and continuous audit (Van Hillo & Weigand, 2016). By recognizing the entangled nature of financial transactions and the instantaneous impact of certain events on financial states, the model can account for the complex interactions within an organization's financial system. The metaphorical use of teleportation emphasizes the need for accounting systems that can rapidly and accurately reflect changes in financial information across different parts of the organization. The model acknowledges the role of entropy in increasing the uncertainty and complexity of accounting for WIP. By applying quantum concepts, the model aims to provide a more nuanced understanding of the uncertainties involved and suggests strategies for managing and reducing entropy in financial information systems[3].

### *Applying the Time-Independent Schrödinger Equation (TISE)*

The Time-Independent Schrödinger Equation (TISE) is a fundamental concept in quantum mechanics (Specogna & Trevisan, 2011) that we can use to explain Quantum WIP; it describes the time evolution of a quantum system. It is given by the equation:

$$\hat{H}\psi = E\psi$$

- $\hat{H}$ represents the Hamiltonian operator of the system
- $\psi$ is the system's wave function
- $E$ is the eigenvalue of the system's total energy

---

[3] AIS keeps systems from decoherence, but with the gradual culling of labor, we can predict a potential collapse due to entropy - unless there is accountancy to ensure systems stay coherent.

The solution to this equation provides information about the behavior of quantum systems. In Work in Progress (WIP) accounting, applying the TISE would involve treating the accounting system as a quantum mechanical system. This would require the identification of the Hamiltonian operator Ĥ that describes the accounting system and its interactions with the environment. The Hamiltonian operator would include terms representing the energy associated with the accounting system's state, such as the energy associated with the accounting entries and their relationships. The potential energy function $U(x)$ would represent the constraints and rules governing the accounting system, such as accounting standards and practices.

Applying the TISE to WIP accounting would allow for calculating the "energy levels" of the accounting system, which could be interpreted as the various states of the accounting system. These energy levels could represent different stages of the accounting process, such as initial recording, adjustments, and final reporting. This quantum accounting model for WIP offers a theoretical framework that captures WIP valuation's dynamic and uncertain nature, providing a foundation for developing more accurate and responsive accounting practices.

The measurement process would involve periodic assessments of WIP, considering the progress of production, cost changes, and other relevant factors that could affect its valuation. This could be facilitated by adopting a probabilistic approach to accounting, where the value of WIP is expressed as a range of possible values with associated probabilities rather than a single deterministic figure. As specific events occur (e.g., completion of a production stage, sale of goods), the probabilistic values can be updated to reflect new information, ultimately leading to a more precise valuation of WIP upon its completion. This methodology, inspired by quantum measurement theory, offers a novel way to address the challenges of WIP accounting, providing a framework better aligned with modern production processes' complexities and uncertainties.

**Discussion**

Traditional WIP accounting methods, such as job costing and process costing, rely on deterministic models (CHENG, 1991) that may not fully capture the complexities and uncertainties of modern business operations. These methods often assume linear progress and clear-cut distinctions between different stages of production, which may not always be the case. In contrast, by incorporating principles like superposition and entanglement, quantum accounting approaches offer a more nuanced and dynamic framework that can adapt to the probabilistic nature of business activities[(Hussain et al., 2011)](#).

Work in Process (WIP) inventory, Finished Goods, and raw materials are interconnected in the manufacturing process and can be considered quantum-interdependent. This interdependence arises from the fact that WIP inventory is not a single, well-defined state but rather a superposition of multiple possible states, each representing a different stage of the production process. The states of raw materials, WIP, and finished goods inventory accounts can affect each other.

The Time-Dependent Schrödinger Equation (TDSE) accounts for how the wavefunction changes with time. It involves both spatial and temporal derivatives of the wavefunction (Schneider & Collins, 2005). On the other hand, the TISE focuses solely on the spatial behavior of the wavefunction. It does not consider time explicitly. Applying this concept to WIP inventory, the "time independence" aspect can be interpreted as focusing on the valuation and state of WIP at a specific point in time without directly accounting for how these values may change over time. In quantum mechanics, the TISE helps identify stable states of a system that are not explicitly time-dependent. Similarly, in accounting for WIP, using a "time-independent" approach could allow for the modeling of WIP valuation as a quantum system in a stable state, representing the inventory's value and condition at a specific moment. In practical terms, the TISE helps us find the allowed energy levels (quantized energies) and corresponding wavefunctions for a given potential energy landscape.

Understanding the quantized energies and wavefunctions the Schrödinger Equation provides could benefit accounting, potentially reshaping accounting standards and practices by adopting quantum theory. Current standards, primarily based on classical accounting principles, may need to be revised to accommodate the probabilistic and interconnected nature of quantum accounting. This could involve the development of new reporting frameworks that allow for the representation of financial data in multiple states and the recognition of the entangled relationships between different financial entities. Additionally, implementing quantum accounting could necessitate changes in auditing practices, with auditors requiring new tools and methodologies to assess the accuracy of quantum-based financial statements.

**Conclusion**

This paper investigated the application of quantum theory to accounting, explicitly focusing on Work in Progress (WIP) valuation. By viewing WIP as existing in a quantum superposition state, we present a novel approach to addressing the uncertainties in valuing partially completed goods. This quantum-inspired perspective aligns with evolving accounting standards that acknowledge the complexity of modern business practices.

Our analysis underscores the value of quantum theory in enriching accounting practices. Employing this approach allows for developing a more sophisticated framework that reflects the intricacies of contemporary business realities. Understanding the quantum perspectives of accounting will allow for more efficient quantum circuits and smoother implementation of quantum technology.

Future research should focus on bridging the gap in understanding the application of quantum theory to accounting, including developing more comprehensive accounting information systems frameworks.

Integrating quantum theory into managerial accounting, particularly in cost accounting, is essential to adapt to the complexities of modern business. Traditional methods are inadequate for accurately capturing the dynamic nature of WIP valuation. Adopting quantum principles offers new insights and tools to enhance accounting practices. This work lays the groundwork for the emerging field of quantum accounting, challenging traditional paradigms and suggesting new directions for research and practice. By embracing quantum principles, accountants can unlock new insights and tools to improve the accuracy and responsiveness of accounting practices. This work aims to lay the foundations for the emerging field of quantum accounting, challenging traditional accounting paradigms and suggesting new avenues for research and practice.

---

**Declaration of Conflicting Interests**

The author declared no potential conflicts of interest with respect to the research, authorship, and/or publication of this article.

**Funding**

The author received no financial support for the research, authorship, and/or publication of this article.

**ORCiD**

https://orcid.org/0009-0007-0352-6330

**Citations:**

http://arxiv.org/abs/2311.11925

Almeida, R., Abrantes, R., Romão, M., & Proença, I. (2020). *The Impact of Uncertainty in the Measurement of Progress in Earned Value Analysis*. 457–467. https://doi.org/10.1016/J.PROCS.2021.01.191

Budd, C. S. (2010). Traditional measures in finance and accounting, problems, literature review, and TOC measures. *Theory of Constraints Handbook*, 335–372.

CHENG, T. C. E. (1991). An Economic Order Quantity Model with Demand-Dependent Unit Production Cost and Imperfect Production Processes. *IIE TRANSACTIONS*. https://doi.org/10.1080/07408179108963838

Davis, M. J., & Heller, E. J. (1981). Quantum dynamical tunneling in bound states. *The Journal of Chemical Physics*, *75*(1), 246–254. https://doi.org/10.1063/1.441832

De Oliveira, K. V., & Lustosa, P. R. B. (2022). The entanglement of accounting goodwill: Einstein's "spooky action at a distance." *Accounting Forum*, 1–22. https://doi.org/10.1080/01559982.2022.2089319

Demski, J., FitzGerald, S. A., Ijiri, Y., Ijiri, Y., & Lin, H. (n.d.). *Quantum Information and Accounting Information:*

Demski, J. S., FitzGerald, S. A., Ijiri, Y., Ijiri, Y., & Lin, H. (2006). Quantum information and accounting information: Their salient features and conceptual applications. *Journal of Accounting and Public Policy*, *25*(4), 435–464. https://doi.org/10.1016/j.jaccpubpol.2006.05.004

Demski, J. S., FitzGerald, S. A., Ijiri, Y., Ijiri, Y., & Lin, H. (2009). Quantum information and accounting information: Exploring conceptual applications of topology. *Journal of Accounting and Public Policy*, *28*(2), 133–147. https://doi.org/10.1016/j.jaccpubpol.2009.01.002

Fellingham, J. C., Lin, H., & Schroeder, D. (2018). Quantum Entropy and Accounting. *SSRN Electronic Journal*. https://doi.org/10.2139/ssrn.3220892

Fellingham, J., Lin, H., & Schroeder, D. (2022). Entropy, Double Entry Accounting and Quantum Entanglement. *Foundations and Trends® in Accounting*, *16*(4), 308–396. https://doi.org/10.1561/1400000069

Fellingham, J., & Schroeder, D. (2006). Quantum information and accounting. *Journal of Engineering and Technology Management*, *23*(1–2), 33–53. https://doi.org/10.1016/j.jengtecman.2006.02.004

Frino, R. A. (2015). Quantum Superposition, Parallel Universes and Time Travel. *viXra*. https://consensus.app/papers/quantum-superposition-parallel-universes-time-travel-frino/ed44c9f2285955fdb31b8a4f108d200c/

Hussain, O., Dillon, T., Hussain, F. K., & Chang, E. (2011). Probabilistic assessment of financial risk in e-business associations. *Simulation Modelling Practice and Theory*, *19*(2), 704–717. https://doi.org/10.1016/j.simpat.2010.10.007

Kovács, G. (2016). LOGISTICS AND PRODUCTION PROCESSES TODAY AND TOMORROW. *Acta Logistica*, *3*(4), 1–5. https://doi.org/10.22306/al.v3i4.71

Lazirko, M. (2023). *Quantum Computing Standards & Accounting Information Systems* (arXiv:2311.11925). arXiv. http://arxiv.org/abs/2311.11925

Lee, R. S. (2021). Quantum finance forecast system with quantum anharmonic oscillator model for quantum price level modeling. *International Advance Journal of Engineering Research*, *4*(02), 01–21.

Myrelid, A., & Olhager, J. (2015). Applying modern accounting techniques in complex manufacturing. *Industrial Management & Data Systems*, *115*(3), 402–418. https://doi.org/10.1108/IMDS-09-2014-0250

Orrell, D. (2019). Quantum Financial Entanglement: The Case of Strategic Default. *SSRN Electronic Journal*. https://doi.org/10.2139/ssrn.3394550

Peres, A. (2000). Classical interventions in quantum systems. I. The measuring process. *Physical Review A*, *61*(2), 022116. https://doi.org/10.1103/PhysRevA.61.022116

Schneider, B. I., & Collins, L. A. (2005). The discrete variable method for the solution of the time-dependent Schrödinger equation. *Journal of Non-Crystalline Solids*, *351*(18), 1551–1558. https://doi.org/10.1016/j.jnoncrysol.2005.03.028

Specogna, R., & Trevisan, F. (2011). A discrete geometric approach to solving time independent Schrödinger equation. *Journal of Computational Physics*, *230*(4), 1370–1381. https://doi.org/10.1016/j.jcp.2010.11.007

Swieringa, R., Gibbins, M., Larsson, L., & Sweeney, J. L. (1976). Experiments in the Heuristics of Human Information Processing. *Journal of Accounting Research*, *14*, 159. https://doi.org/10.2307/2490450

Van Hillo, R., & Weigand, H. (2016). Continuous Auditing & Continuous Monitoring: Continuous value? *2016 IEEE Tenth International Conference on Research Challenges in Information Science (RCIS)*, 1–11. https://doi.org/10.1109/RCIS.2016.7549279

Wang, X.-L., Cai, X.-D., Su, Z.-E., Chen, M.-C., Wu, D., Li, L., Liu, N.-L., Lu, C.-Y., & Pan, J.-W. (2015). Quantum teleportation of multiple degrees of freedom of a single photon. *Nature*, *518*(7540), 516–519. https://doi.org/10.1038/nature14246

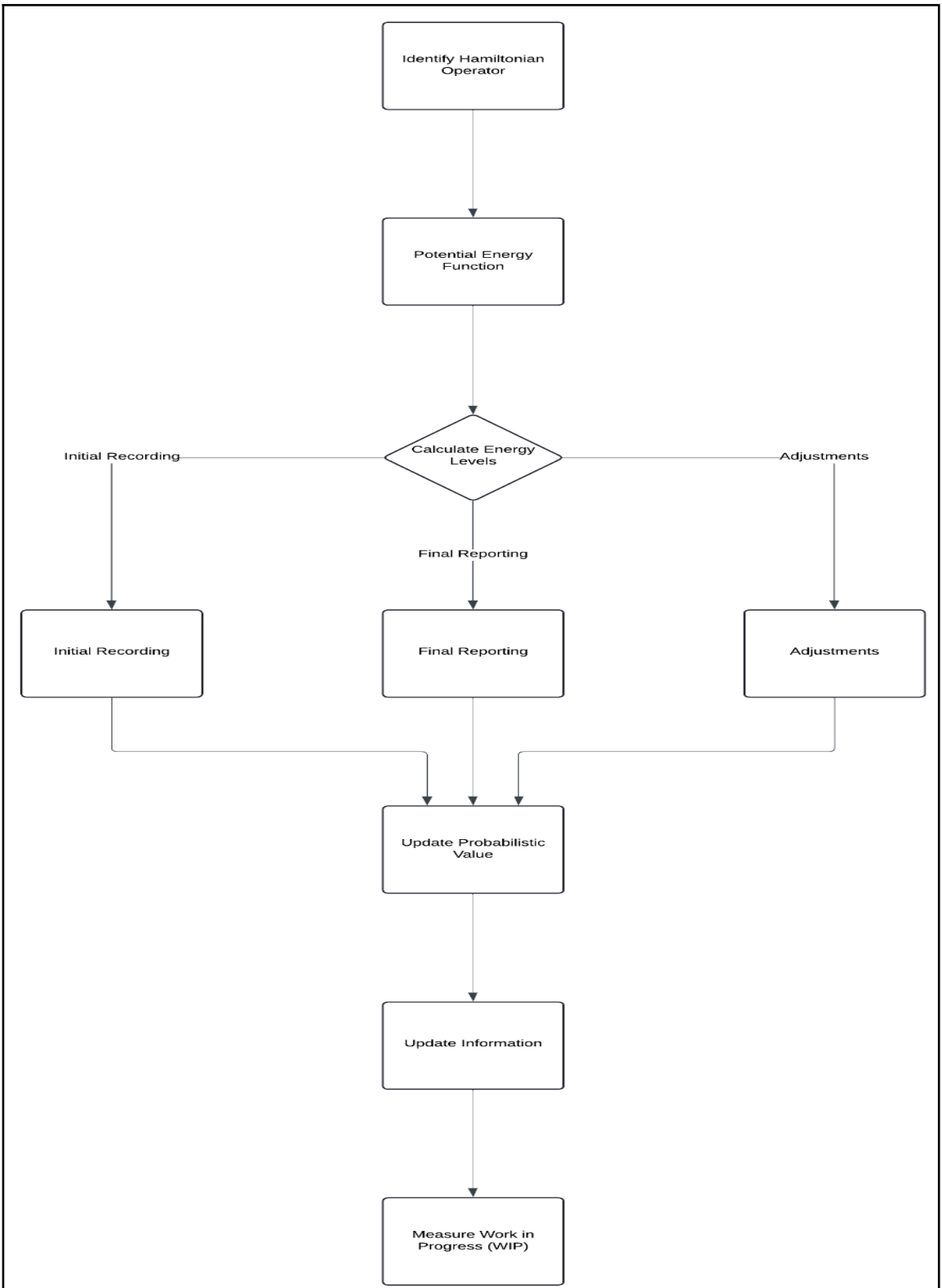

**Graphical Abstract**

**APPENDIX**

In the context of quantum WIP accounting, the particle-in-a-box model illustrates a scenario where a quantum system, representing WIP, is confined within a limited space delineated by impenetrable barriers. These barriers can represent the constraints and boundaries of the WIP account, including project timelines, resource availability, and regulatory requirements.

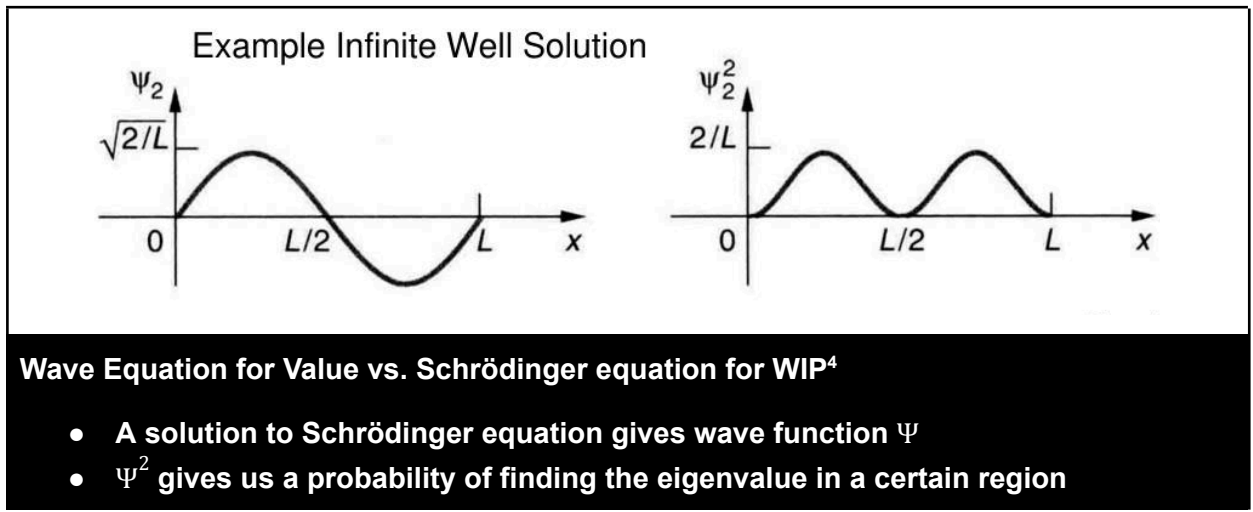

**Wave Equation for Value vs. Schrödinger equation for WIP[4]**
- **A solution to Schrödinger equation gives wave function $\Psi$**
- **$\Psi^2$ gives us a probability of finding the eigenvalue in a certain region**

Much like the particle in a box model in quantum mechanics, where the behavior of the particle is influenced by the size and shape of the confinement, the dynamics of WIP within the accounting system are shaped by various internal and external factors.

---

[4] IMAGE ADAPTED FROM
https://www.slideserve.com/hayfa-david/topic-5-schr-dinger-equation